\documentclass[]{article}
\usepackage{amsmath,amssymb,bbm}
\usepackage{geometry}
\usepackage{authblk}
\usepackage{setspace}
\usepackage[document]{ragged2e}
\usepackage{xcolor}
\usepackage[superscript]{cite}
\usepackage{graphicx}
\usepackage{subcaption}
\usepackage{algorithm}
\usepackage{algpseudocode}

\newcommand{\SI}[1]{SI~{\color{blue} #1}}

\title{Quantum-enhanced magnetometry by phase estimation algorithms with a
	single artificial atom}

\author[1]{S.\ Danilin}
\author[2,4]{A.V.\ Lebedev}
\author[1]{A.\ Veps\"al\"ainen}
\author[3,4]{G.B.\ Lesovik}
\author[2]{G.\ Blatter}
\author[1]{G.S.\ Paraoanu}
\affil[1]{Low Temperature Laboratory, Department of Applied Physics, Aalto University School of Science, PO Box 15100, Aalto FI-00076, Finland}
\affil[2]{Theoretische Physik, Wolfgang-Pauli-Strasse 27, ETH Z\"urich, CH-8093 Z\"urich, Switzerland}
\affil[3]{L.D.\ Landau Institute for Theoretical Physics RAS, Akad.\ Semenova av., 1-A, Chernogolovka, 142432, Moscow Region, Russia;}
\affil[4]{Moscow Institute of Physics and Technology, Institutskii per.\ 9,
	Dolgoprudny, 141700, Moscow District, Russia}

\date{}

\begin{document}
\begin{titlepage}
\maketitle	

\thispagestyle{empty}

Corresponding authors:

A.V.\ Lebedev, postal address, telephone, lebedev@itp.phys.ethz.ch,

G.S.\ Paraoanu, postal address, telephone, sorin.paraoanu@aalto.fi
\end{titlepage}	

{\bf Abstract}

Phase estimation algorithms are key protocols in quantum information
processing. Besides applications in quantum computing, they can also be
employed in metrology as they allow for fast extraction of information stored in the quantum state of a system. Here, we implement two suitably modified phase estimation procedures, the Kitaev- and the semiclassical
Fourier-transform algorithms, using an artificial atom realized with a
superconducting transmon circuit. We demonstrate that both algorithms yield a flux sensitivity exceeding the classical shot-noise limit of the device, allowing one to approach the Heisenberg limit.  Our experiment paves the way for the use of superconducting qubits as metrological devices which are potentially able to outperform the best existing flux sensors with a sensitivity enhanced by few orders of magnitude.


{\bf Introduction}

Phase estimation algorithms are building elements for many important quantum algorithms\cite{Mosca:1998}, such as Shor's factorization
algorithm\cite{Shor:1994,Smolin:2013} or Lloyd's algorithm\cite{Lloyd:2009}
for solving systems of linear equations. At the same time, phase estimation is a natural concept in quantum metrology\cite{Giovanetti:2004}, where one aims at evaluating an unknown parameter $\lambda$ that typically enters into the Hamiltonian of a probe quantum system and defines its energy states
$E_n(\lambda)$. In a standard (classical) measurement, the precision
$\delta\lambda$ is restricted by the shot-noise limit $\delta \lambda \propto 1/\sqrt{t}$, where $t$ is the measurement time.  This, however, is not a fundamental limit: in principle, the ultimate attainable precision scales as $\delta \lambda \propto 1/t$, constrained only by the Heisenberg relation $\Delta E(\lambda) \geq 2\pi\hbar/t$, where $\Delta E = \max_{n,m} (E_n - E_m)$. The Heisenberg limit can be achieved with the help of entanglement resources, e.g., using NOON photon states in optics\cite{Dowling:2008,Nagata:2007, Matthews:2016}. However, these states are difficult to create in general and they typically have a short coherence time. Alternatively, one can reach the Heisenberg limit without exploiting entanglement, by using the coherence of the wavefunction of a single quantum system as a dynamical resource. However, the uncontrollable interaction of the probe with the environment limits the time scale $t$ where the Heisenberg scaling can be attained by the probe's coherence time $t\sim T_2$. A further improvement then has to make use of an alternative measurement strategy with a precision following the standard quantum limit but with a better prefactor.

The unknown parameter $\lambda$ can be estimated from the phase $\phi = \Delta E(\lambda) \, \tau/\hbar$ accumulated by the system in the course of its evolution during the time $\tau \sim T_2$.  The $2\pi$-periodicity of the phase limits the probe's measurement range $\Delta\lambda$ where $\lambda$ can be unambiguously resolved within the narrow interval
$[\delta\lambda]_\mathrm{H} = 2\pi\hbar/(\mu T_2)$, with $\mu \equiv \partial \Delta E/\partial \lambda$ denoting the sensitivity of the probe's spectrum. Therefore, the improvement in the precision at larger $T_2$ is concomitant with a proportional reduction of the measurement range $\Delta\lambda$.  The use of phase estimation algorithms then allows to resolve the $2\pi$ phase uncertainty and hence break this unfavorable trade-off between the measurement precision $\delta\lambda$ and the measurement range $\Delta\lambda$. Moreover, a metrological procedure based on a phase estimation algorithm is Heisenberg-limited: it attains the resolution $\delta\lambda \sim [\delta\lambda]_\mathrm{H}$ within a large measurement range $\Delta\lambda \gg [\delta\lambda]_\mathrm{H}$ with a Heisenberg scaling in the phase accumulation time $\tau$, i.e., $\delta \lambda \propto \hbar/(\mu \tau)$ for $\tau \leq T_2$.  At larger times $\tau > T_2$, the measurement proceeds with independent measurements involving the optimal time delay $\tau = T_2$. Running $N = t/T_2$ experiments and averaging over $N\gg 1$ outcomes, one can further improve the precision within the standard quantum limit\cite{Sekatski:2017}, $\delta \lambda \propto 2\pi\hbar/(\mu T_2 \sqrt{N}) \equiv  2\pi\hbar/(\mu \sqrt{t T_2})$.

There are two major classes of phase estimation algorithms, one suggested
early on by Kitaev\cite{kitaev:1995} and a second originating from the quantum Fourier transform\cite{Nielsen:2004,Dam:2007}.  In quantum computing, the Kitaev algorithm was run as part of Shor's factorization
algorithm\cite{Martinez:2016} and the Fourier transform algorithm was used in optics to measure frequencies\cite{Higgins:2007}. These algorithms are
system-independent and can be employed in a variety of experimental
settings, e.g., using NV centers in diamond for the sensitive detection of
magnetic fields\cite{Jelezko:2012,Wrachtrup:2014,Bonato:2016}.

{\bf Results}

Here, we implement a modified version of these algorithms using an artificial atom or qubit in the form of a superconducting transmon circuit\cite{Koch}. We show that the transmon can be operated as a dc flux magnetometer with Heisenberg-limited sensitivity. The sensitivity is boosted by a magnetic moment that is about five orders of magnitude larger than that of natural atoms or ions. The idea of the experiment is to combine the extreme magnetic-field sensitivity of superconducting quantum interference devices (SQUIDs) with an enhanced performance brought about by exploiting quantum coherence.   The `quantum' in the name of this device refers to the
macroscopic complex wave function of the superconducting electronic state. In the SQUID loop geometry, the relative phase of the superconducting
wavefunctions across the Josephson junctions acquires a dependence on magnetic flux $\Phi$ via the Aharonov-Bohm effect. However, despite its quantum origin, in standard SQUID measurements this phase is a classical variable. In contrast, for the SQUID loop of a transmon qubit, the phase turns into a fully dynamical quantum observable and the flux $\Phi$ dependence is encoded in the energy-level separation $\hbar \omega_{01} (\Phi)$ between the ground state and the first excited state.  Therefore, it is possible to exploit the phase difference $\phi = [\omega_\mathrm{d} - \omega_{01}(\Phi)] \tau$ acquired during a time $\tau$ by the qubit when it is prepared into a coherent superposition of the ground and excited energy states and driven by an external microwave field at a frequency $\omega_\mathrm{d}$. Differently from their ``natural'' counterparts, where the characteristics of the quantum sensor are sample independent and defined by the atomic structure, for artificial-atom systems, such as the transmon, we need to adapt the algorithms by including device-specific properties in a so-called \textit{passport} -- a sample specific Ramsey interference pattern obtained in advance from characterization measurements, see Fig.~1b. Making use of phase estimation algorithms, we demonstrate an enhanced dc-flux sensitivity of the transmon sensor in an enlarged flux range as compared to standard (classical) measurement schemes. Recently, a standard measurement procedure using a flux qubit has been used for the measurement of an ac-magnetic field signal\cite{Lupascu:2012}.
\begin{figure}
\includegraphics[width=\textwidth]{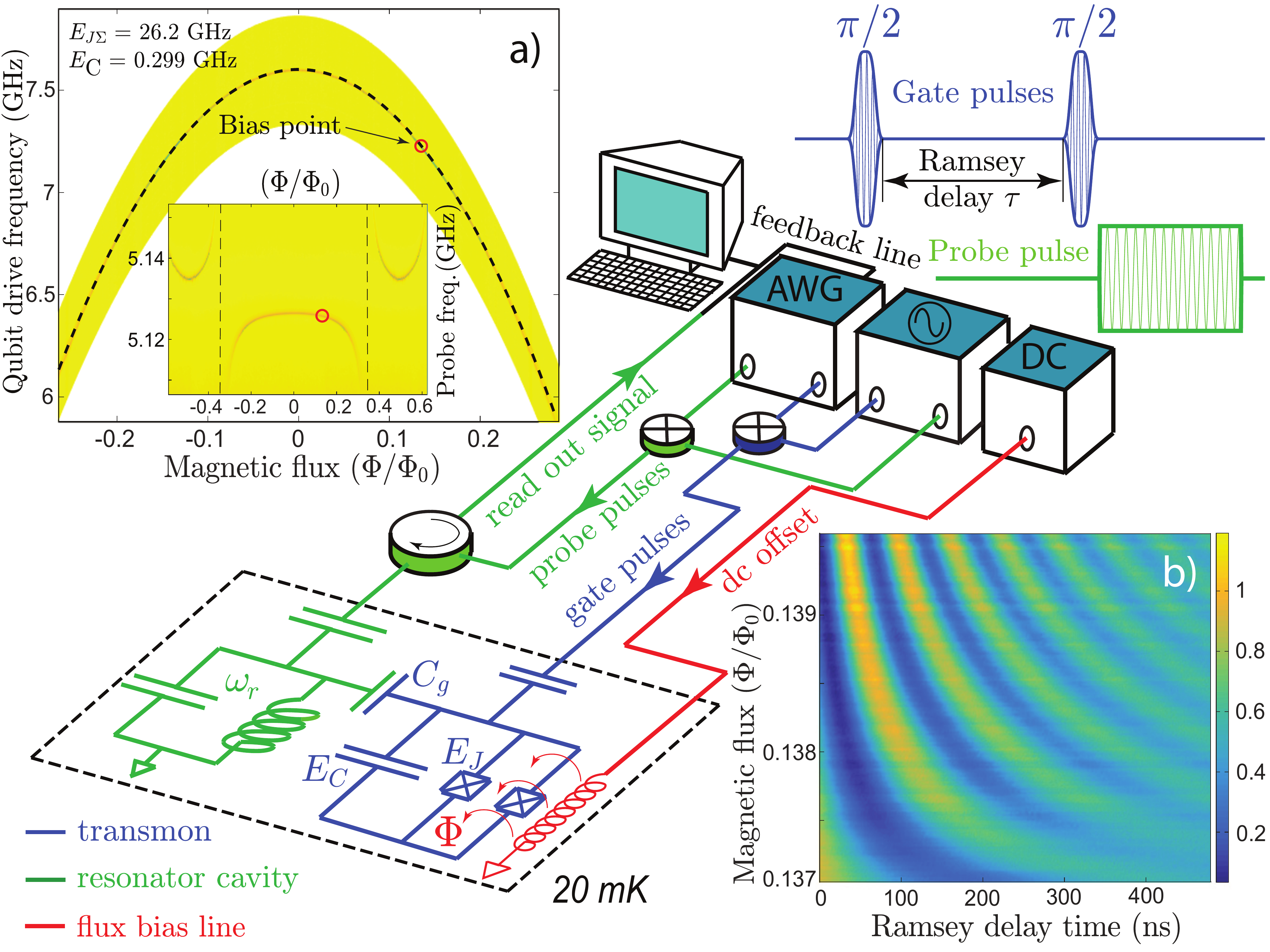}
\caption{\textbf{Experimental layout}. The schematic shows a transmon qubit (in blue) comprised of a capacitor and a SQUID loop with two nearly identical junctions. The qubit charging and total Josephson energies are $E_\mathrm{C} = 299\ \textrm{MHz}$ and $E_\mathrm{J\Sigma} =26.2$ GHz. The qubit is coupled via a gate capacitor $C_\mathrm{g}$ to a coplanar waveguide resonator (CPW, in green) with a resonance frequency $\omega_\mathrm{r}$  around $2\pi\ \times\ 5.12\ \textrm{GHz}$. The magnetic flux $\Phi$ through the transmon's SQUID loop is controlled by a dc-current flowing through a flux-bias line (in red). An arbitrary waveform generator (AWG) and a microwave analog signal generator are employed to create a Ramsey sequence of two $\pi/2$ microwave pulses at a carrier frequency $\omega_\mathrm{d} = 2\pi\ \times\ 7.246\ \textrm{GHz}$ separated by a time delay $\tau$. The sequence drives the transmon into a superposition of ground and excited states where the state amplitudes depend on the accumulated phase $\phi = [ \omega_\mathrm{d}-\omega_{01}(\Phi)] \tau$. The qubit state is read out nondestructively using a probe pulse sent to the CPW resonator; the reflected signal is downconverted (not shown in the figure), digitized, and analyzed by a computer. Next, the computer updates a flux distribution function $\mathcal{P}(\Phi)$ stored in its memory, determines the next optimal Ramsey delay time, and feeds it back into the AWG. \textbf{a)} Qubit transition frequency $\omega_{01}(\Phi)$ as a function of magnetic flux $\Phi$ (parabolic curve). The bottom inset shows the CPW resonator's spectrum.  The red circles indicate the bias point of our transmon
sensor: we operate far away from the 'sweet spot' in a regime where the
transmon's frequency $\omega_{01}(\Phi)$ is an approximately linear function of the flux $\Phi$ within the entire flux range $\Delta\Phi$. For the fluxes around the point considered here, the frequency $\omega_\mathrm{r}$ of the
readout CPW resonator remains approximately constant. \textbf{b)} A
pre-measured sample-specific Ramsey interference fringes pattern defines the 'passport' function of our sensor. This can be regarded as a non-normalized
probability function $P_p(\tau,\Phi)$ to observe the qubit in the excited
state after a Ramsey sequence with a delay $\tau$ for a specific value of the magnetic flux $\Phi$. The largest flux value used to obtain the Ramsey
interference fringes pattern $\Phi = 0.1394\ \Phi_0$ corresponds to a frequency detuning $\Delta\omega = \omega_\mathrm{d} - \omega_{01}(\Phi) =
2\pi\times15.8\ \textrm{MHz}$ between the drive and the qubit transition
frequencies. The flux range of the 'passport' $\Delta\Phi \sim 2.5\times
10^{-3}\ \Phi_0$ corresponds to a range $2\pi\times 13.8\ \textrm{MHz}$ in
frequency detuning. }
\end{figure}

The experiment employs a superconducting circuit in a transmon configuration, consisting of a capacitively-shunted split Cooper-pair box coupled to a $\lambda/4$\---wavelength coplanar waveguide (CPW) resonator realized in a 90 nm thick aluminum film deposited on the surface of a silicon substrate, see Fig.\ 1 and SI 1 for an image of the sensor device. The SQUID loop of the transmon has an area of $S\simeq 600\ \mathrm{\mu m}^2$, which is chosen large in comparison with standard transmon qubit designs in order to provide a higher sensitivity to magnetic-field changes.  The magnetic moment of this artificial atom is $\mu = S \hbar\, |d\omega_{01}/d \Phi|$, directly proportional to the area $S$ and the rate of change with flux $\Phi$ of the transition frequency $\omega_{01}$. For our device, we obtain $d\omega_{01}/d\Phi = -\ 2\pi\ \times\ 5.3\ {\rm GHz}/ \Phi_0$ at the bias point, resulting in $\mu = 1.10 \times 10^{5}\ \mu_{\rm\scriptscriptstyle B}$, where $\mu _{\rm\scriptscriptstyle B}$ is the Bohr magneton. By comparison, the Zeeman splitting due to the magnetic moment of NV centers is $28\ \textrm{GHz}\ \textrm{T}^{-1}$, corresponding to a magnetic moment of $2\mu_{\rm\scriptscriptstyle B}$. The sample is thermally anchored to the mixing chamber plate of a dilution refrigerator and cooled down to a temperature of roughly $\sim 20\ \textrm{mK}$. The qubit has a separate flux-bias line and a microwave gate line, the former allowing to change the qubit transition frequency, while the latter is used for the qubit's state manipulation. The qubit state is determined by a nondemolition read-out technique (see Methods and \SI 1) measuring the probe signal reflected back from the dispersively-coupled CPW resonator. To increase the magnetic field sensitivity, we bias the qubit away from the 'sweet spot', see Fig.\ 1a. This follows an opposite strategy as compared to the situation where the phase estimation algorithms are employed for quantum computing and simulations; in the latter cases, the qubit sensitivity to flux noise is maximally suppressed by tuning the device to the `sweet spot'
characterized by a vanishing first derivative of the energy with respect to
flux. Operating away from the 'sweet spot' leads to a reduction of the $T_2$ time. The decoherence rate $T_2^{-1}= (2T_1)^{-1} + T_\phi^{-1}$ is the sum of the relaxation $(2T_1)^{-1}$ and dephasing $T_\phi^{-1}$ rates\cite{Clarke:2008}. The dephasing rate appreciably increases at our bias point, which reduces $T_2$ and thus the number of available steps that can be implemented in the Kitaev- and Fourier algorithms.

In the experiment, we apply a Ramsey sequence of two consecutive $\pi/2$
pulses separated by a time delay $\tau$, which corresponds to an effective
spin-$1/2$ precession around the z-axis of the Bloch sphere. The precession
angle $\phi = \Delta\omega(\Phi) \tau$ is defined by the frequency mismatch
$\Delta\omega(\Phi) = \omega_\mathrm{d}-\omega_{01}(\Phi)$ between the
transition frequency $\omega_{01}(\Phi)$ of the transmon qubit and the fixed drive frequency $\omega_\mathrm{d}$ of the $\pi/2$ pulses. The Ramsey sequence drives the transmon from its ground state into a coherent superposition of ground- and excited states with relative amplitudes determined by the phase $\phi$. The theoretical probability to find the transmon in the first excited state is given by
\begin{equation}
P\bigl[\tau,\Delta\omega(\Phi)\bigr] = \frac12 + \frac12 \exp(-\tau/
2T_1) \, \gamma(\tau) \cos\bigl[\Delta\omega(\Phi)\tau\bigr]
\label{eq:p1}
\end{equation}
and depends both on the delay time $\tau$ and on the magnetic flux $\Phi$
through the frequency mismatch $\Delta\omega (\Phi)$. The decay function
$\gamma(\tau)$ accounts for qubit dephasing, typically due to charge or flux noise.  By design, the transmon artificial atom is rather insensitive to background charge fluctuations. On the other hand, intrinsic $1/f$ magnetic-flux noise couples to the SQUID loop and is known to be a relevant source for dephasing in flux qubits\cite{Yoshihara:2006,Bialczak:2007,Anton:2013} ;  in addition, other decoherence mechanisms can be present, see below for details. The
dephasing process can be described through an external classical noise source, see Methods. The particular shape of $\gamma(\tau)$ depends on the noise spectral density at low frequencies. 'White' noise with a constant power density results in an exponential decay function $\gamma(\tau) =
\exp(-\Gamma_\mathrm{wn}\tau)$, while $1/f$-noise produces a Gaussian decay
$\gamma(\tau) = \exp\bigl[-(\Gamma_{1/f}\tau)^2\bigr]$.  We fit our
experimental curves $P(\tau,\Delta\omega(\Phi))$ by Eq.\ (\ref{eq:p1}) using both an exponential and a Gaussian decay, see SI 2. For our sample with a
relaxation time $T_1$ of about $260$ ns, we cannot distinguish between these two fits, neither in the 'sweet spot' nor in the bias point. Fitting the
Ramsey oscillation at different fluxes one finds $\Gamma_\mathrm{wn}^{-1}
\approx 1250$ ns and $\Gamma_{1/f}^{-1} \approx 780$ ns at the 'sweet
spot'. At the bias point, these pure dephasing times reduce to $520$ ns and
$420$ ns, respectively. The decay rates $\Gamma_\mathrm{wn}$ and
$\Gamma_{1/f}$ in the bias point then can be translated into
equivalent white and $1/f$ flux noises and we find the spectral densities
$S_\mathrm{wn} = (5.9 \times 10^{-8}\ \Phi_0)^2/\mathrm{Hz}$ and $S_{1/f}(f)= (1.9 \times 10^{-5}\ \Phi_0)^2/f[\mathrm{Hz}]$, respectively (see Methods).

The function $\gamma(\tau)$ determines the optimal delay time $\tau$ where the sensitivity of the probability $P(\tau,\Delta\omega)$ to the changes in
$\Delta\omega$  and hence to a flux is the highest.  In the standard
(classical) measurement approach, a minimal delay $\tau=\tau_0 \ll T_2$ sets the frequency range $\Delta\omega(\Phi) \in [0,\pi/\tau_0]$ where the phase
$\phi$ and hence $P(\tau,\Delta\omega)$ can be unambiguously resolved. This
defines the range $\Delta\Phi = \pi (\tau_0 d\omega_{01}/d\Phi)^{-1}$ where
the magnetic flux can be resolved with a precision scaling given by the
standard quantum limit (see Methods),
\begin{equation}
[\delta\Phi]_\mathrm{class} = \Bigl| \frac{d\omega_{01}(\Phi)}{d\Phi} \Bigr|^{-1}
\frac1{\tau_0\sqrt{t/T_\mathrm{rep}}} = \frac{A_\mathrm{class}}{\sqrt{t}},
\label{eq:cl_precision}
\end{equation}
where $t$ is the total measurement or \textit{sensing} time of the experiment and $T_\mathrm{rep}$ is the time duration of a single Ramsey measurement. A better flux sensitivity can be attained at larger delays $\tau$, where the probability $P[\tau,\Delta\omega(\Phi)]$ is more sensitive to changes in $\Delta\omega$. We obtain the best sensitivity at $\tau = \tau^*$ defined by the condition $(2T_1)^{-1} - [\ln\gamma(\tau)]' = \tau^{-1}$ (see Methods),
\begin{equation}
[\delta\Phi]_\mathrm{quant} = \Bigl| \frac{d\omega_{01}(\Phi)}{d\Phi} \Bigr|^{-1}
\frac{e}{\tau^*\sqrt{t/T_\mathrm{rep}}} =  \frac{A_\mathrm{quant}}{\sqrt{t}}.
\label{eq:q_precision}
\end{equation}
The amplitudes $A_\mathrm{class}$ and $A_\mathrm{quant}$ in Eqs.
(\ref{eq:cl_precision}) and (\ref{eq:q_precision}) quantify the magnetic flux sensitivities. Measuring at the optimal delay $\tau = \tau^*$ improves the flux resolution by a factor $A_\mathrm{class}/A_\mathrm{quant}=
\tau^*/(e\tau_0)$, which depends on the qubit's coherence time, the latter
serving as the quantum resource in our algorithms. Another important factor
which enhances the flux sensitivity is the slope $d\omega_{01}/d\Phi$ of the transmon's spectrum. At our working point $\omega_{01} = 2\pi\ \times\ 7.246$ GHz, we have $d\omega_{01}/d\Phi \ = -\ 2\pi\ \times\ 5.3~\rm{GHz}/\Phi_0$. The minimal delay is given by $\tau_0 \approx 31.6$ ns, see SI 1.  The repetition time $T_\mathrm{rep} = 6.546~\mu$s involves the maximal time duration of the Ramsey sequence, the duration of the probe pulse ($2~\mu$s) and the transmon's relaxation time back into its ground state ($4~\mu$s, which is 15 times longer than the $T_1$ time). Combining these numbers and setting $\tau^* \sim 2T_1$, we estimate the theoretical value of flux sensitivity for our transmon sensor as $A_\mathrm{quant}\simeq 4\times 10^{-7}\ \Phi_0\,\mathrm{Hz}^{-1/2}$, see Eq.~(\ref{eq:q_precision}), providing an improvement by a factor $A_\mathrm{class} / A_\mathrm{quant} \sim 6$ over the classical sensitivity. Note, that the best sensitivity is attained at $T_\mathrm{rep} = \tau^*$ (i.e., for a very fast control and readout) that gives for our sample $[\delta\Phi]_\mathrm{quant} \simeq 1.1 \times 10^{-7}\
\Phi_0\,\mathrm{Hz}^{-1/2}/\sqrt{t}$.

Measuring at large time delays $\tau \sim T_2$ leaves an uncertainty in
$\Delta\omega(\Phi)$ due to the multiple $2\pi$-winding of the accumulated
phase, thereby squeezing the flux range $\Delta\Phi \sim 2.5\times10^{-3}
\Phi_0$ by the small factor $\tau_0/T_2$. The Kitaev- and Fourier phase
estimation algorithms, avoid this phase uncertainty by measuring the
probability $P(\tau,\Delta\omega)$ at different delays $\tau_k = 2^k \tau_0$ for $K \sim \log_2(T_2/\tau_0)$ consecutive steps $k =0,\dots, K-1$. As a
result, such a metrological procedure is able to resolve the magnetic flux
with the quantum limited resolution $[\delta\Phi]_\mathrm{quant}$, see
Eq.~(\ref{eq:q_precision}), within the original flux range $\Delta\Phi$ set by the duration $\sim \tau_0$ of the control rf-pulses. The operation of the Kitaev and Fourier metrological procedures can be viewed as a successive
determination of the binary digits of the index $n = [b_{K-1}\dots b_0] \equiv \sum_{k=0}^{K-1} b_k\,2^k$ in the so-called quantum abacus\cite{Suslov:2011}. The Kitaev algorithm starts from a minimal delay $\tau = \tau_0$ and determines the most significant bit $b_{K-1}$ in its first step, further proceeding with the less significant bits $b_{K-2},\dots, b_0$.  The Fourier algorithm works backwards\cite{Griffiths:1996}: it starts from the maximal
delay $\tau\sim T_2$ and first determines the least significant bit $b_0$,
then gradually learns more and more significant bits $b_1, b_2,\dots,
b_{K-1}$.

\textbf{Modified Kitaev- and Fourier metrological algorithms.} In the present work, we use modified versions of the phase estimation protocols, which take into account the nonidealities present in actual experiments. For brevity, we will still refer to these protocols as the Kitaev- and Fourier phase estimation algorithms. We demonstrate the superiority of these algorithms over the standard technique and show that we can beat the standard quantum limit. Instead of relying on the ideal theoretical probability function $P[\tau,\Delta\omega(\Phi)]$ of Eq.\ (\ref{eq:p1}) these modified Kitaev and Fourier protocols exploit the empirical probability $P_p(\tau,\Phi)$, the so-called passport, which we measure by a set of Ramsey sequences at various magnetic fluxes $\Phi$, representing the result on a discrete equidistant grid in the form $P_p(\tau_j,\Phi_i)$, see Fig.\ 1b. Here, $\Phi_i = (i-1) [\delta\Phi]_\mathrm{step} + \Phi_1$ with the index $i$ chosen from the flux-index set $I_0 =[1,161]$, $[\delta\Phi]_\mathrm{step} \simeq 1.59 \times 10^{-5}\ \Phi_0$, $\Phi_1 \simeq 0.137\ \Phi_0$, and discrete time delays $\tau_j = (j-1)\times 2$ ns, $j =1, \dots 241$ quantifying the time separation between the two $\pi/2$ rf-pulses of the Ramsey sequence. In order to increase the signal-to-noise ratio, we average over $65000$ Ramsey experiments at each discrete point $(\tau_j, \Phi_i)$. The resulting pattern is only approximately
described by Eq.\ (\ref{eq:p1}) due to the fact that the resonator frequency changes slightly with the applied flux, thus modifying our calibration (see
Methods and \SI 2). In principle, one can change the working point to an even more sensitive part of the spectrum  at the price of a further distortion of this pattern.

Using the qubit passport $P_p(\tau_j,\Phi_i)$, one can pose the following
metrological question: given an unknown flux $\Phi$ within some pre-chosen
range, how can one estimate its value using a minimal number of Ramsey
measurements? We design two metrological algorithms where the time delay
$\tau$ of the Ramsey sequence serves as an adaptive parameter whose value is dynamically adjusted. In the course of operation, both our algorithms return a discrete probability distribution ${\cal P}(\Phi_i)$, $i\in I_0$, which
reflects our current knowledge about the flux $\Phi$ to be measured. This probability distribution is improved in subsequent steps and shrinks to a
narrow interval around the actual flux-value when running the algorithm.

{\bf Bayesian learning.} The elementary building block for both our
metrological algorithms is a Bayesian learning subroutine which updates the
discrete flux distribution ${\cal P}(\Phi_i)$ after each Ramsey measurement of the qubit state. This subroutine takes the time delay $\tau_j$ between $\pi/2$ pulses as an input parameter and performs a sequence of $N=32$ Ramsey measurements.  Our readout scheme returns a measured variable $h_N$ which, at $N\gg 1$, is equal to the empirical passport probability $P_p(\tau_j,\Phi_i)$. At small values $N$, the readout variable $h_N$ is a normally distributed random variable  with a mean value given by $P_p(\tau_j,\Phi_i)$,
\begin{equation}
p(h_N|\tau_j,\Phi_i) = \frac1{\sqrt{2\pi}\sigma_{N}}
\exp\left[-\frac{(h_N-P_p(\tau_j,\Phi_i))^2}{2\sigma_{N}^2}\right],
\end{equation}
where the variance $\sigma_{N}^2 = \sigma_1^2/N$ can be directly measured,
$\sigma_1^2 \approx 3.5$ (see \SI 1 for further explanations on the readout
variable $h_N$). Next, the algorithm makes use of the measurement outcomes
$h_N$ and updates the flux probability distribution with the help of Bayes'
rule, ${\cal P}(\Phi_i) \to p(h_N|\tau_j,\Phi_i)\,{\cal P}(\Phi_i)/\sum_i
p(h_N|\tau_j,\Phi_i)\,{\cal P}(\Phi_i)$.

{\bf Kitaev algorithm.} The Kitaev-type metrological algorithm has been
introduced earlier in Ref.\ \cite{Lebedev:2014}. The algorithm involves
$K$ steps $k = 0, \dots, K-1$ with optimized Ramsey times $\tau_k$, tolerances $\epsilon_k$, and flux index sets $I_k$; below, ${\cal N}(I)$ denotes the size of a discrete set $I$. It is initialized with a uniform discrete distribution ${\cal P}_0(\Phi_i)$ which reflects our prior ignorance of the flux to be measured.  In the first step $k=0$, the algorithm repeats the Bayesian learning subroutine at a zero time delay $\tau^{(0)} = 0$ between $\pi/2$ pulses until the probability distribution shrinks to a twice narrower interval $I_1 \subset I_0$, i.e., ${\cal N}(I_1) = {\cal N}(I_0)/2$, satisfying $\sum_{i\in I_1} {\cal P}_0(\Phi_i) \geq 1-\epsilon_0$. The flux values $\Phi_i, i\notin I_1$ are discarded. After completing the first step, the algorithm searches for the optimal delay $\tau_j$ for the next step. The next optimal Ramsey measurement requires a larger delay $\tau^{(1)}> 0$ such that the passport $P_p(\tau^{(1)},\Phi_i)$, $i\in I_1$, has the largest range:
$\tau^{(1)} = \mathrm{argmax}_{\tau_j} \Delta P(\tau_j)$ where $\Delta
P(\tau_j) = \max_{i\in I_1} P_p(\tau_j,\Phi_i) - \min_{i\in I_1}
P_p(\tau_j,\Phi_i)$. The algorithm thus sweeps over the passport data
$P_p(\tau_j,\Phi_i)$ to find the optimal delay $\tau^{(1)}$ with maximal range $\Delta P(\tau^{(1)})$.  Subsequently, a new distribution ${\cal
P}_1(\Phi_{i\in I_1}) = {\cal N}^{-1}(I_1)$ and ${\cal P}_1(\Phi_{i\notin
I_1}) = 0$ is initialized and the algorithm proceeds to the next step by
running the Bayesian learning with the new optimal delay $\tau_1$. After $K$ steps, the algorithm localizes $\Phi$ within a $2^K$ times narrower interval $I_K$, ${\cal N}(I_K) = {\cal N}(I_0)/2^K$, with an error probability
$\epsilon = 1- \prod_{k=0}^{K-1} (1-\epsilon_k)$.

{\bf Quantum Fourier algorithm.} This algorithm starts from the Ramsey
measurement with an optimal time delay $\tau^{(s)} \sim T_2$. The starting
delay $\tau^{(s)}$ is a free input parameter of the algorithm. Similarly to
the Kitaev algorithm, the quantum Fourier algorithm runs the Bayesian learning subroutine until the flux probability distribution ${\cal P}_0(\Phi_i)$, $i\in I_0$, squeezes to a twice narrower subset $S_1 \subset I_0$ such that $\sum_{i\in S_1} {\cal P}_0(\Phi_i) \geq 1-\epsilon_0$. However, in contrast to the Kitaev algorithm, the passport function $P_p(\tau^{(s)},\Phi_i)$ is an ambiguous function of $\Phi_i$ at the large delay $\tau^{(s)}$. As a result, $S_1$ is not a single interval but rather a set of $n \sim \tau^{(s)}/\tau_0$ disjoint narrow intervals $S_1 = I_1 \cup \dots \cup I_n$ of almost equal lengths, see Fig.\ 2.  Hence, after completing the first step, the flux value is distributed among $n$ equiprobable alternatives $I_i$. The Fourier algorithm discriminates between these $n$ alternatives in the next steps. First, it searches for the next optimal delay $\tau_j$, where it is possible to rule out half of the remaining alternatives in the most efficient way. At each delay $\tau_j$ the algorithm splits the remaining intervals $I_i$, $i=1,\dots, n$ into two approximately equal-in-size groups $A=I_{i_1} \cup \dots \cup I_{i_{[n/2]}}$ and $B = I_{i_{[n/2]+1}} \cup \dots \cup I_{i_n}$ which are ordered by the passport function, $P_p(\tau_j, \Phi\in A) > P_p(\tau_j,\Phi\in B)$.  Then it finds the probability distance $\Delta P(\tau_j) = \min_{i\in A} P_p(\tau_j,\Phi_i) - \max_{i\in B} P_p(\tau_j,\Phi_i) >0$ separating the two sets $A$ and $B$. Repeating this procedure at all available delays $\tau_j$, the algorithm finds the optimal delay $\tau^{(1)}$  with maximal $\Delta P(\tau_j)$  over the discrete set of delays $\tau_j$.  In the next step, the algorithm discriminates between $A$ and $B$ by repeating the Bayesian learning subroutine approximately $[\Delta P(\tau^{(1)})]^{-2}$ times and sets $S_2 = A$ or $B$. Continuing in this way, the algorithm returns a single interval $I_\mathrm{out}$ where the actual value of the flux $\Phi(\Phi_i)$, $i\in I_\mathrm{out}$, is located. Fig.\ 2 shows how the flux distribution function ${\cal P}(\Phi_i)$ develops in time during the execution of the Kitaev and Fourier algorithms.
\begin{figure}
	\includegraphics[width=\textwidth]{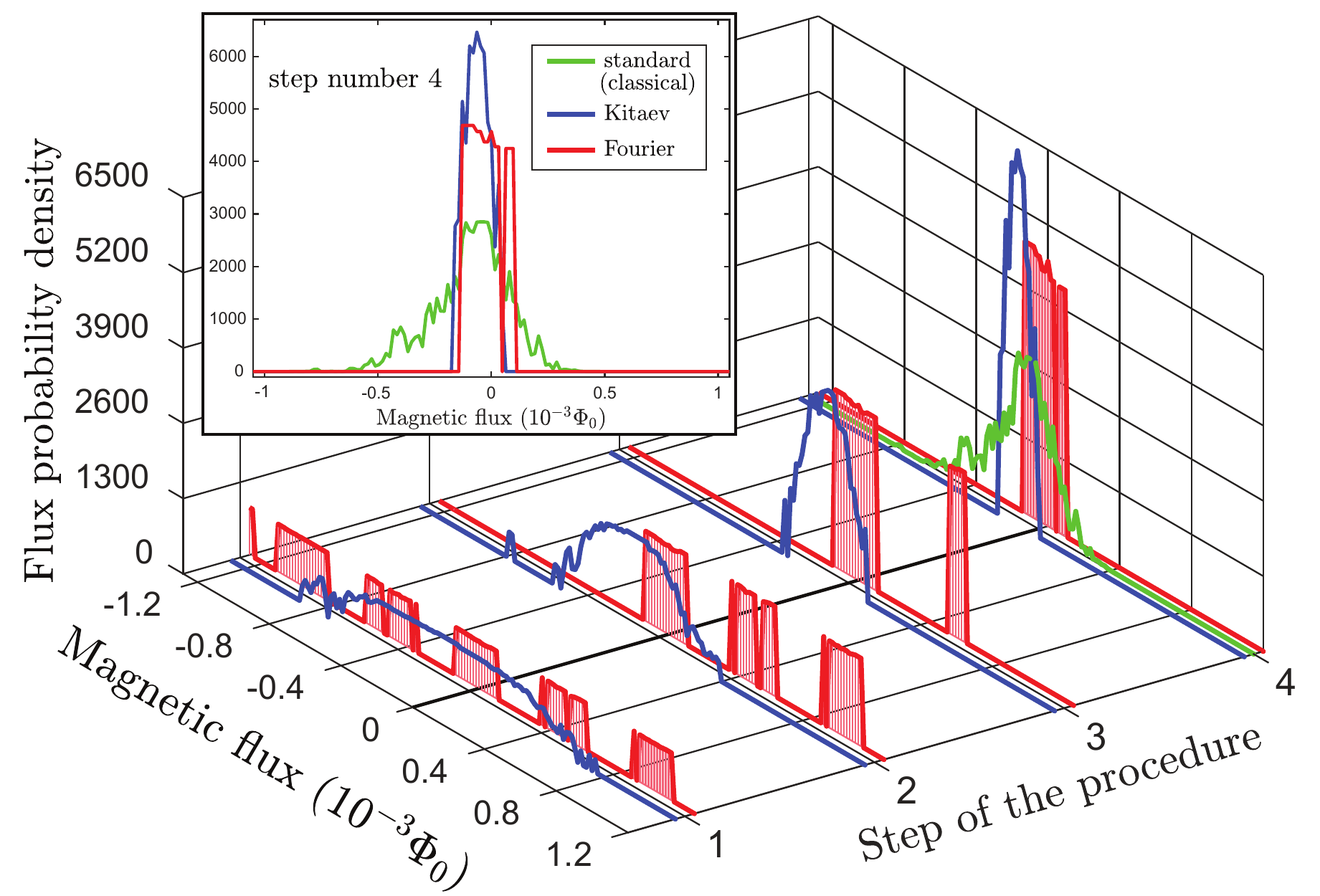}
	\caption{Evolution of the flux probability distribution ${\cal P}_k( \Phi_i - \Phi)$, during the run of the first $k = 1, \dots, 4$ steps of the Fourier (red panels) and Kitaev (blue curves) estimation algorithms. The magnetic flux is measured with respect to a reference flux (as explained in the Methods). The actual flux value is shown by a thick black line in the flux-step plane. The Kitaev algorithm starts at a zero delay $\tau = 0$ and a first step returns a broad probability distribution with a single peak centered near the actual flux value. During the run of the Kitaev algorithm this peak narrows down. The Fourier algorithm starts from the Ramsey measurement at large delay $\tau^{(s)}=360$ ns, with the first step returning a probability distribution with six out of twelve flux intervals assuming a non-vanishing value. Hence, this first step selects half of the $n=\tau^{(s)}/\tau_0 \sim 12$ different flux intervals $\Delta\Phi_m$ given by $\Delta\omega_m = \Delta\omega_0 + 2\pi i/\tau_s$, $m=0,\dots,n-1$, where $\Delta\omega_m \equiv \omega_{01}(\Delta\Phi_m) - \omega_\mathrm{d}$ is the frequency interval corresponding to the flux interval $\Delta\Phi_m$, and determines the parity of the yet unknown index $m\in[0,n -1]$ associated with the true flux interval. In the second step, the Fourier algorithm proceeds to a shorter delay and rules out another half of the remaining six intervals. In the next two steps the algorithm discriminates between the remaining three alternatives and ends up with the correct flux interval. The green line at the fourth step displays the probability distribution learned by the standard (classical) procedure during the same number of Ramsey measurements as was required by the quantum procedures. The distributions obtained at the step number 4 for the Kitaev and Fourier estimation algorithms and in the standard (classical) measurement are shown in the inset. }
\end{figure}

\textbf{Results.} The superiority of our quantum metrological algorithms is
clearly demonstrated by the scaling behaviour of the magnetic flux resolution with the total \textit{sensing time} of the flux measurement, see Fig.\ 3. We run each algorithm $n = 25$ times at every flux value
$\Phi=\Phi_i$, $i\in I_0$, within the entire flux range, and find the
corresponding arrays of estimated values $\hat\Phi_{ji}$, $j=1,\dots, n$. The estimate $\hat\Phi = \hat\Phi\bigl({\cal P}(\Phi)\bigl)$ is defined as the most likely value derived from the observed probability distribution ${\cal 	P}(\Phi)$. For a probability distribution ${\cal P}_i(\Phi)$ measured at a known flux value $\Phi = \Phi_i$ the corresponding estimate $\hat\Phi_{ji}$ is a random quantity due to statistical nature of the measurement procedure. We define an \textit{aggregated resolution} $\delta \Phi$ as an ensemble standard deviation of the random variables $\hat\Phi_{ji}-\Phi_i$,
\begin{equation}
\delta \Phi^2 = \frac1{{\cal N}(I_0)}\sum_{i\in I_0} \frac1{n-1}
\sum_{j=1}^{n} \bigl[ \hat\Phi_{ji} - \Phi_i\bigr]^2.
\label{eq:accuracy}
\end{equation}
In case of the Fourier algorithm, such a definition is meaningful only at the final step of the algorithm where ${\cal P}(\Phi)$ becomes a single-peaked function. The sensing time $t$ is defined through the total number of calls of the Bayesian learning subroutine $m$, $t = N T_\mathrm{rep}\, m$. The scaling behaviour of the measured flux resolution $\delta\Phi(t)$ with sensing time $t$ is shown in Fig.\ 3 for both our algorithms and is compared with the scaling $\delta \Phi_\mathrm{std}(t)$ of the standard (classical) procedure, where all Ramsey measurements are done at a zero delay $\tau = 0$.
\begin{figure} \includegraphics[width=\textwidth]{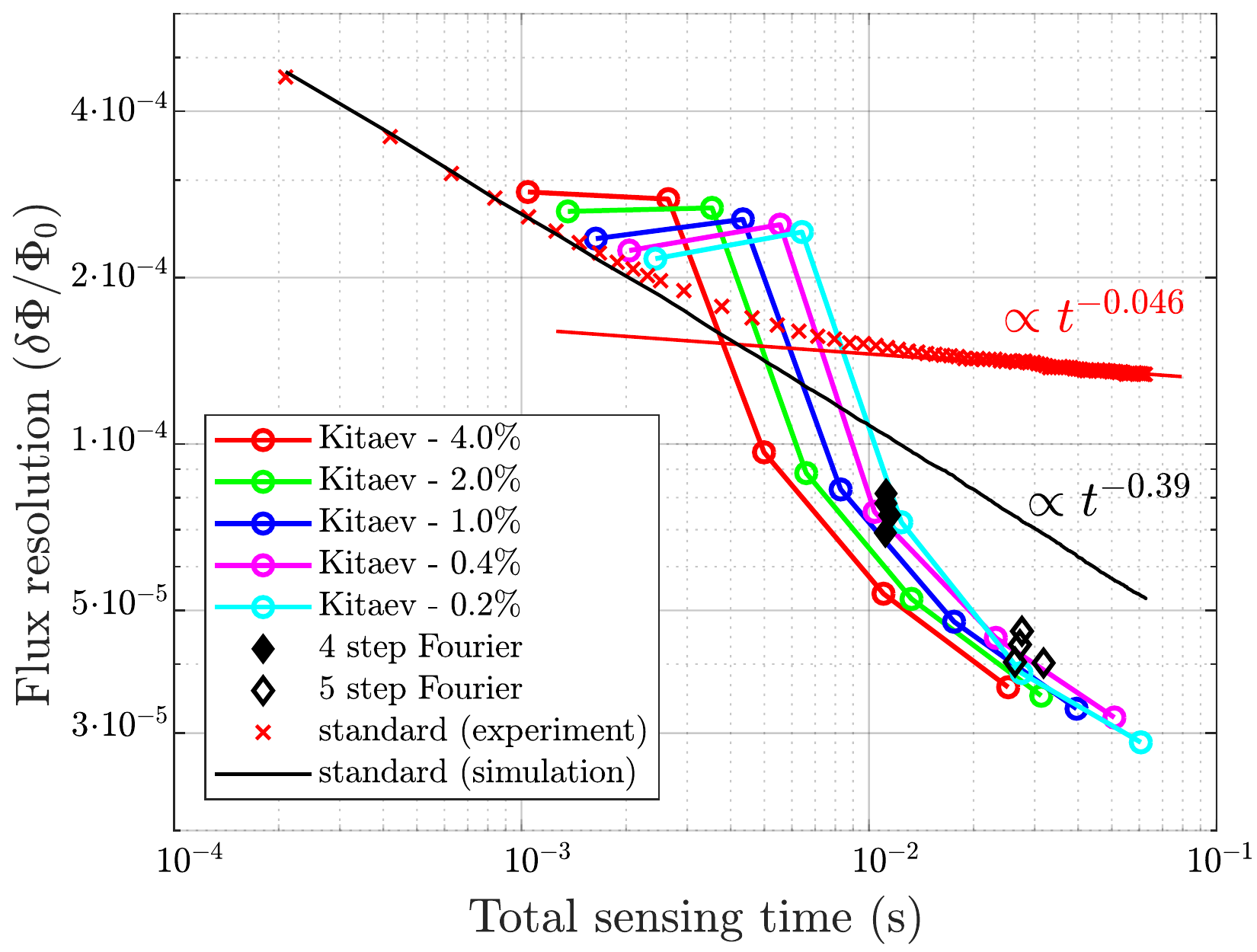}
	\caption{Observed scaling behavior of the flux resolution
versus total sensing time for the three different metrological
procedures, Kitaev (colored circles), Fourier (black diamonds) and
standard (red crosses). The Kitaev algortihm has been run with
constant tolerances $\epsilon_k = \epsilon$ for each step
$k=1,\dots,5$ and for five different values of $\epsilon$ as indicated
by different colors.  The Fourier algorithm has been performed with
the step-dependent tolerances $\epsilon_k = 0.182,\, 0.076,\, 0.039,\,
0.02,\, 0.01$ for $k=1,\dots, 5$.  We show the result of the Fourier
algorithm only for the final two steps, $k=4$ (filled diamonds) and
$k=5$ (empty diamonds), running the algorithm with four different
starting delays, $\tau^{(s)} = 300,\, 320,\, 340$ and $360$ ns
(all collapsed to the same data points).  The phase estimation
algorithms lag behind in precision at short times when compared to the
standard procedure, but rapidly gain precision at longer times.
The black solid line represents the scaling law for a numerical
simulation of the standard procedure with a regular passport function
given by  Eq.\ (\ref{eq:p1}). The crossover to the red solid
line is due to the irregularity of the passport function. }
\end{figure}

Both quantum algorithms clearly outperform the standard procedure, with the
Kitaev algorithm appearing slightly more efficient than the Fourier one.  We explain this by the fact that the Fourier algorithm strongly relies on the
periodicity of the Ramsey interference pattern given by Eq. (\ref{eq:p1}),
whereas our readout scheme produces a slightly distorted pattern. On the other hand, the Kitaev algorithm turns out to be more stable to the irregularities in the measured passport function $P_p(\tau,\Phi)$.  The magnetic flux sensitivities $A_\mathrm{quant}$ range within $5.6 - 7.1\times 10^{-6}\ \Phi_0\, \mathrm{Hz}^{-1/2}$ for the Kitaev algorithm and within $6.5-8.5 \times 10^{-6}\ \Phi_0\,\mathrm{Hz}^{-1/2}$ for the Fourier procedure.  These sensitivites are an order of magnitude worse than the theoretical bound $4.0\ \times\ 10^{-7}\ \Phi_0\,\mathrm{Hz}^{-1/2}$ set by Eq.\ (\ref{eq:q_precision}). The discrepancy has two main reasons.  First, our readout scheme is not a single-shot measurement, which leads to a factor $32$ increase of the $T_\mathrm{rep}$ time. Second, we spend part of the time resource for the intermediate steps with $\tau_k \leq T_2$ during the run of the phase estimation procedure. Finally, for our transmon, the SQUID area $S \simeq \ 20 \times 30\ \mathrm{\mu m}^2$ results in a magnetic filed sensitivity in the range $19.3 - 29.3\ \mathrm{pT}\,\mathrm{Hz}^{-1/2}$.

Decoherence processes define the most important factor limiting the
sensitivity of our device. E.g., the intrinsic $1/f$ flux
noise\cite{Bialczak:2007,Anton:2013} caused by magnetic impurities constitutes a relevant source of decoherence.  At short times $0 < \tau < 2T_1$, $\tau$ the duration of a single Ramsey sequence, the presence of $1/f$ noise can be accounted for by a finite coherence time $T_\phi$ of the qubit.  Assuming that dephasing originates exclusively from intrinsic flux noise results in an upper limit $S_{1/f}(f) =(1.9 \times 10^{-5}\ \Phi_0)^2/f[\mathrm{Hz}]$ for the noise spectral function. At much larger time scales, as defined by the entire duration $t$ of the Kitaev or Fourier procedure, $1/f$ noise causes low-frequency flux fluctuations $\langle \delta\Phi^2\rangle \sim \int_{1/t}^{1/\tau^*} \!\! S_{1/f}(f)\, df$. As follows from Fig.\ 3, the 5-step Kitaev procedure takes $\approx 0.05$ s, which provides a value $\langle \delta\Phi^2 \rangle \sim (6.4 \times 10^{-5}\ \Phi_0)^2$ for the flux fluctuations, about twice larger than the actually achieved flux resolution $\delta\Phi \sim 3\times 10^{-5}\ \Phi_0$. This suggests that $1/f$ flux noise has a smaller weight and another, non-magnetic decoherence mechanism is present in our device. One of the potential candidates derives from electron tunneling at defects inside the dielectric layer of the qubit's Josephson junctions. These fluctuating charges produce $1/f$ noise in the critical current and hence affect the transition frequency of the transmon atom\cite{SQUID:2006}. The $1/f$ flux noise may become more pronounced at a larger size $L$ of the transmon's SQUID loop as the flux-noise spectral density grows linearly with the loop size\cite{Bialczak:2007}. Consequently, the flux resolution $[\delta\Phi]$ degrades $\propto \sqrt{L}$ when increasing the loop size, while the corresponding magnetic field resolution $[\delta B] \propto [\delta \Phi]/L^2$ still improves as $L$ growths, see Methods.

Interestingly, the non-ideality of the qubit's passport strongly affects the performance of the standard procedure as well. Its scaling behaviour $\delta \Phi_\mathrm{std} \propto t^{-\alpha}$ exhibits a crossover in the scaling
exponent $\alpha$, assuming a value $\alpha \approx 0.39$ at short sensing
times, while at large times $\alpha$ decreases to a much smaller value
$\approx 0.046$.  The scaling exponent $0.39$ deviates from the shot noise
exponent $1/2$ due to the cases when the actual flux value is located near the boundaries of the flux interval $\Phi\in[\Phi_1,\Phi_{161}]$, where the
passport function $P_p(0,\Phi)$ has an extremum and the scaling exponent for the standard procedure is reduced to $1/4$, $\delta \Phi(t)\propto t^{-1/4}$. As a result, the aggregated scaling exponent of Eq.\ (\ref{eq:accuracy}) is reduced below $1/2$.  On the other hand, the crossover to $\alpha \approx 0.046$ is a consequence of the irregularity of the passport function set by low-frequency noise fluctuations during the passport measurement.  Indeed, at large sensing times, the standard procedure needs to distinguish fluxes within a narrow interval where the passport function $P_p(\tau=0,\Phi)$ has a non-regular and non-monotonic dependence on $\Phi$. As a result, the Bayesian learning procedure fails to converge to a correct flux value. In contrast, at a larger scale of $\Phi$, the passport function is smooth and monotonic and the standard procedure behaves properly. These arguments are indeed confirmed by a numerical simulation with a regular passport function given by Eq.\ (\ref{eq:p1}). Importantly, both our quantum metrological algorithms are more stable than the standard procedure with respect to passport imperfections and their scaling behaviour at large sensing times coincides with the scaling
behaviour resulting from a regular passport function. The quantum algorithms suffer, however, from the same irregularity problem at larger sensing times, not shown in Fig.\ 3.

Finally, we discuss how our metrological algorithms use the quantum resource of qubit coherence in order to acquire information about the measured flux. We quantify the quantum coherence resource spent in a given measurement by the total \textit{phase accumulation time} $\tau_\phi = N\sum_k \tau_k m_k$, where $m_k$ is the number of calls for the Bayesian learning subroutine with delay $\tau_k$. The amount of information $\Delta I$ acquired during the measurement is given by a decrease of the Shannon entropy $\Delta I = H({\cal P}_0) - H({\cal P})$, where ${\cal P}_0(\Phi_i)$ and ${\cal P}(\Phi_i)$ are the initial and final probability distributions and $H({\cal P}) = -\sum_i {\cal 	P}(\Phi_i) \log_2({\cal P}(\Phi_i))$. The scaling behaviour $\Delta I(\tau_\phi) \propto \tau_\phi^\alpha$ separates the classical domain with $0<\alpha \leq 0.5$ from the quantum domain with $0.5< \alpha \leq 1$, where $\alpha = 1$ corresponds to the ultimate Heisenberg limit. Indeed, in the ideal case where no relaxation and decoherence phenomena are present, the quantum algorithms double the flux resolution (squeeze the flux distribution function into a twice narrower range) for each next step of the procedure. This means that the associated Shannon entropy decreases by $\ln(2)$ and one learns one bit of information for each doubling of the Ramsey delay time. In contrast, the classical procedure with $N \gg 1$ repetitions results in a Gaussian probability distribution of the measured quantity where the precision scales as $\delta\Phi\propto 1/\sqrt{N}$. Hence the associated Shannon entropy scales as $\tau^{0.5}$ with an invested total phase accumulation time $\tau = N\tau_0$. We run each quantum algorithm $25$ times for every flux value and average over the obtained information gains and phase times. The resulting scaling dependence is shown in Fig.\ 4 and demonstrates that both Kitaev and Fourier algorithms indeed belong to the quantum domain with a scaling exponent within $[0.624,0.654]$ ($95$\% confidence interval). The scaling exponent is still below the Heisenberg limit, which is a consequence of the finite dephasing time $T_2$: at large time delays $\tau \sim T_2$ the visibility of the Ramsey interference pattern decreases, requiring more Ramsey measurements in order to learn the next bit of information.
\begin{figure}
	\includegraphics[width=\textwidth]{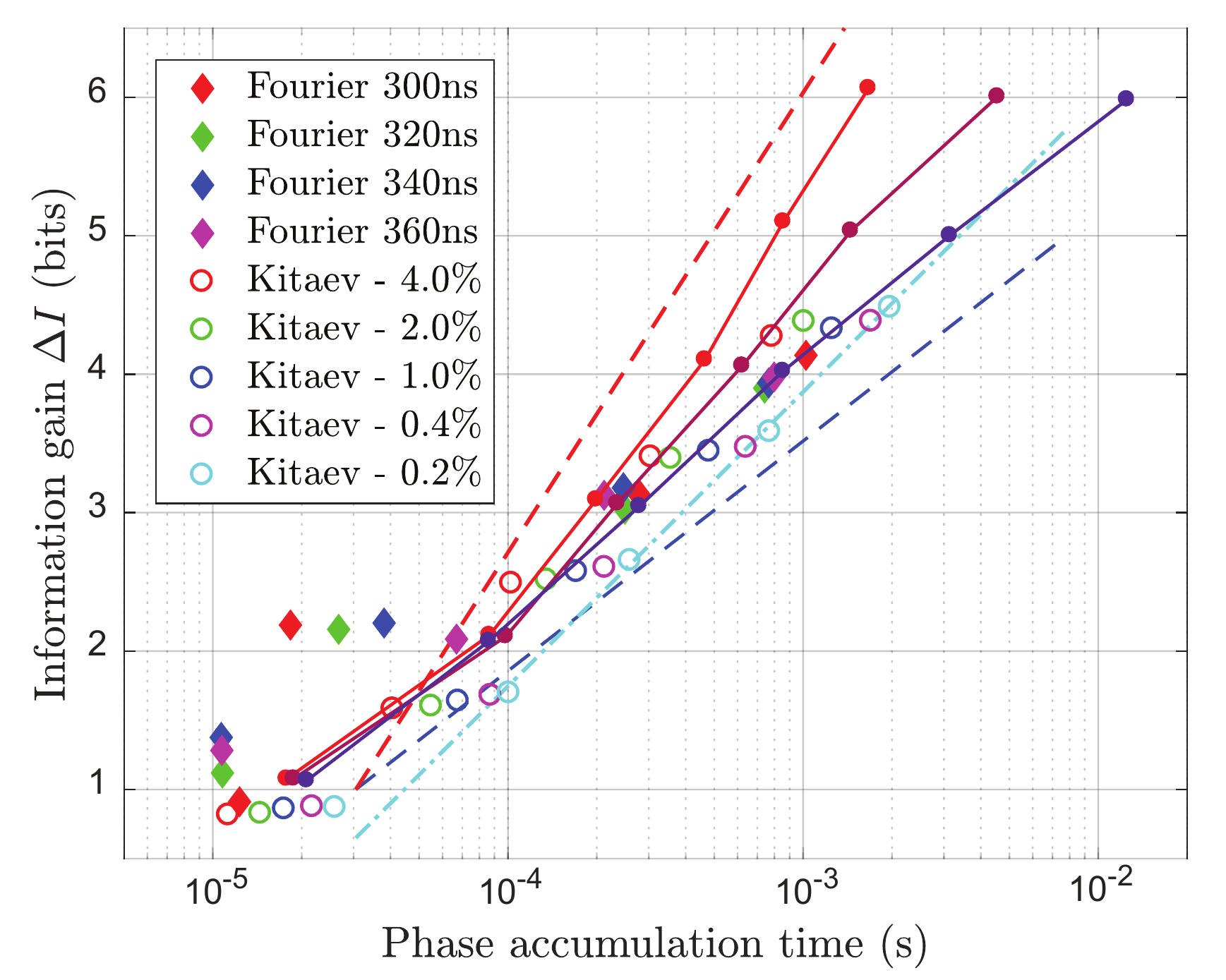}
	\caption{Information (in bits) inferred by the Kitaev (circles) and
Fourier (filled diamonds) algorithms as a function of the total phase
accumulation time. The Kitaev algorithm was run for five different
tolerance level constants $\epsilon$ at each step (indicated by
color). The color of diamonds indicates the different starting time of
the Fourier algorithm. The dashed red and blue lines refer to the
Heisenberg and shot-noise scaling laws with the corresponding scaling
exponents $1$ and $1/2$. The thin solid lines show the numerical
simulation for the 6-step Kitaev algorithm with an idealized passport
function given by Eq.\ (\ref{eq:p1}) at different dephasing times
$T_2$ ranging from $10$ $\mu$s (red line) to $340$ ns (blue line). One
can clearly see that at large dephasing times the Kitaev procedure
approaches the Heisenberg limit, while at smaller $T_2$ the scaling
exponent decreases to the standard quantum limit $0.5$. The observed
experimental scaling behaviour shows that both Fourier and Kitaev
algorithm are indeed quantum with a scaling exponent above the
standard quantum limit $1/2$, see the dash-dotted cyan line connecting
the Kitaev (at 0.2\% tolerance) data.
 }
\end{figure}

{\bf Discussion}

We have used a single transmon qubit as a magnetic-flux sensor and have implemented two quantum metrological algorithms in order to push the measurement sensitivity beyond the standard shot-noise limit. In our
experiments, we utilize the coherent dynamics of the qubit as a quantum resource. We demonstrate experimentally, on the same sensor sample, that suitably modified Kitaev- and Fourier algorithms both outperform the classical shot-noise-limited measurement procedure and approach the Heisenberg limit. Both algorithms exhibit a similar asymptotic flux-sensitivity $A_\Phi\sim 6\times 10^{-6}\ \Phi_0\,\mathrm{Hz}^{-1/2}$ or magnetic-field sensitivity $A_B\sim 20.7\ \mathrm{pT}\,\mathrm{Hz}^{-1/2}$  within \textit{a dynamical 	range $\Delta\Phi/A_\Phi \sim  417\sqrt{\mathrm{Hz}}$} at a coherence time $T_2 \sim 260$ ns of the qubit.

Finally, we can compare the characteristics of our qubit sensor with other
magnetometers. dc-SQUID sensors typically feature a $1\ \mu\Phi_0\,
\mathrm{Hz}^{-1/2}$ sensitivity and a much larger dynamical range $\sim 10^6 \sqrt{\mathrm{Hz}}$, see ref.\ \cite{SQUID:2006}.  However, conventional
dc-SQUIDs are operated with a current bias close to critical, which limits
their sensitivity to $10^{-8}-10^{-6}\ \Phi_0\, \mathrm{Hz}^{-1/2}$, see refs.\ \cite{Awschalom:1988,Wellstood:1989,SQUID:2006}, due to intrinsic
thermal noise fluctuations of excited quasi-particles.  Atomic
magnetometers~\cite{Budker:2007} can approach a magnetic field sensitivity
$\sim 0.1-1.0\  \mathrm{fT}\, \mathrm{ Hz}^{-1/2}$. However, these
magnetometers measure the field in a finite macroscopic volume $\sim 1\
\mathrm{cm}^3$ and their sensitivity translated to the $(100\  \mu
\mathrm{m})^3$ volume range of a transmon sensor reduces to $0.1-1.0\
\mathrm{pT}\, \mathrm{Hz}^{-1/2}$, with a dynamical range $\sim 10^4-10^5
\sqrt{\mathrm{Hz}}$ compatible with dc-SQUID sensors. NV centers in diamond
are able to resolve magnetic fields with atomic spatial resolution and
approach sensitivities $\sim 6.1\ \mathrm{nT} \, \mathrm{Hz}^{-1/2}$. Phase
estimation algorithms allow one to enlarge the dynamical range of
NV-sensors\cite{Jelezko:2012,Wrachtrup:2014,Bonato:2016} up to $3\times 10^5\ \mathrm{Hz}^{1/2}$. With magnetic-field sensors based on superconducting qubits there is a lot of potential for improvements in dynamic range and sensitivity.  In contrast to dc-SQUIDs, such sensors are not prone to thermal noise fluctuations. Their sensitivity is limited only by their coherence time and the duration of the readout procedure. With a coherence time of $T_2\sim 5\ \mu$s and very fast control and readout ($T_{\textrm{rep}}\simeq\tau^\ast\simeq T_2$), one can potentially
access a sensitivity of $A_{\textrm{quant}}\simeq 4\times 10^{-8}\
\Phi_0\, \textrm{Hz}^{-1/2}$ and a dynamical range of
$\Delta\Phi/A_{\textrm{quant}}\simeq\ 6.3\times 10^4\sqrt{\textrm{Hz}}$.
Moreover, making use of the higher excitation levels in a transmon atom, one can increase the sensitivity even further\cite{Shlyakhov:2018}.
\vspace{1cm}

{\bf Methods}

{\bf The superconducting artificial atom}

The transmon \cite{Koch} is a capacitively-shunted split Cooper-pair box, with a Hamiltonian 
\begin{equation}
H = 4 E_{\rm C} n^2 - E_{J}(\Phi) \cos (\varphi),
\end{equation}
where $E_{\rm C}$ is the charging energy $E_{\rm C} = e^2/2 C_{\Sigma}$ with $C_{\Sigma}$ the total capacitance (dominated by the shunting capacitor).  The SQUID loop in the transmon design provides a flux-dependent effective
Josephson energy $E_{J}(\Phi) = E_{\rm J\Sigma} |\cos (\pi \Phi/\Phi_{0})|$
(assuming identical junctions). The state of the device is described by a
wavefunction which treats the superconducting relative phase across junctions $\varphi$ as a quantum variable similar to a standard coordinate.  In contrast to standard SQUID measurements, the flux dependence is reflected in the quantized energy levels; for the first transition this reads
\begin{equation}
\hbar \omega_{01} = \sqrt{8E_{\rm C}E_{\rm J}(\Phi)} - E_{\rm C}.
\end{equation}
The readout of the qubit state is realized by a dispersive coupling of the
transmon to a coplanar waveguide (CPW) resonator whose resonance frequency
depends on the transmon state. This allows us to perform a non-demolition
measurement of the qubit state by sending a probe pulse to the CPW right after the second $\pi/2$-pulse and collecting the resulting resonator response signal whose shape in time depends on the qubit state, see \SI 1.

{\bf Dephasing mechanisms}

The dephasing of the qubit can be modeled via an interaction of the qubit with an external classical noise source $\nu(t)$. The qubit state acquires a stochastic relative phase $\delta\phi = \int^\tau dt \, \nu(t)\, \partial
\omega_{01}/\partial\nu$. Then the decay function $\gamma(\tau) \equiv \langle e^{i\delta\phi}\rangle$ can be expressed via a noise spectral density function $S_\nu(\omega) = \int dtdt^\prime \langle\langle \nu(t) \nu(t^\prime) \rangle\rangle e^{i\omega(t-t^\prime)}$ as $\gamma(\tau) =
\exp \bigl[-\frac12 (\partial \omega_{01}/\partial\nu)^2 \int \frac
{d\omega}{2\pi} S_\nu(\omega) \sin^2(\omega\tau/2) /(\omega/2)^2\bigr]$, see ref. \cite{Cottet:2002}. A white noise source with a constant spectral density $S_\nu=S_\mathrm{wn}$ at low frequencies gives an exponential decay function $\gamma(\tau) = \exp(-\Gamma_\mathrm{wn} \tau)$ with $\Gamma_\mathrm{wn} = \frac12 S_\mathrm{wn} (\partial \omega_{01}/\partial\nu)^2$. A $1/f$-noise $S_\nu(\omega) = S_{1/f}/|\omega|$ gives a Gaussian decay $\gamma(\tau) = \exp[-(\Gamma_\mathrm{1/f}\tau)^2]$ with $\Gamma_\mathrm{1/f} \sim \sqrt{S_{1/f} |\ln(\omega_c\tau)|/(2\pi)} |\partial\omega_{01}/\partial\nu|$, where $\tau \sim 2T_1$ and $\omega_c\sim 1 s^{-1}$ is a low frequency cut-off. One can estimate the corresponding
decay rates $\Gamma_\mathrm{wn} = (520 \mathrm{ns})^{-1}$ and $\Gamma_\mathrm{1/f} = (420 \mathrm{ns})^{-1}$, from the free-induction decay of the qubit state at the working point of the qubit spectrum, see \SI 2. If we assume that the main dephasing mechanism is due to the intrinsic magnetic flux noise of the SQUID loop $\nu(t) = \delta \Phi(t)$, one can translate these rates into the corresponding noise spectral densities, $S_\mathrm{wn}\approx (5.9 \times 10^{-8}\ \Phi_0)^2/\mathrm{Hz}$ and $S_{1/f}(f) \approx (1.9 \times 10^{-5}\ \Phi_0)^2/f[\mathrm{Hz}]$, where we have used a value  $d\omega_{01}/d\Phi \approx -2\pi \times 5.3$ GHz$/\Phi_0$ obtained from the characterization measurement of the qubit spectrum.

{\bf Quantum and classical magnetic flux sensitivities}

After $N$ Ramsey experiments at a fixed delay $\tau$, the probability of the excited state $P(\tau,\Delta\omega)$ can be estimated as $N_1/N$, where $N_1$ is the number of outcomes where an excited state was detected. The accuracy $\delta P^2 = \langle \bigl( P(\tau,\Delta\omega) - N_1/N)^2\rangle$ of this estimate is given by a binomial statistics, $\delta P^2 = N_1(N-N_1)/N^3 \leq 1/(4N)$. From the equation $P(\tau,\Delta\omega) = N_1/N$, one can find the frequency mismatch $\Delta\omega$. The corresponding accuracy $\delta[\Delta\omega]$ can be found from the relation $\delta P = \bigl|\frac{\partial P(\tau,\Delta\omega)}{\partial \Delta\omega}\bigr| \delta[\Delta\omega]$, hence $\delta[\Delta\omega] = \bigl|\frac{\partial 	P(\tau,\Delta\omega)}{\partial \Delta\omega}\bigr|^{-1} \frac1{2\sqrt{N}}$. From Eq.\ (\ref{eq:p1}), it follows that $\textrm{min}_{\Delta \omega}\left(\bigl|\frac{\partial P(\tau,\Delta\omega)}{\partial \Delta\omega}\bigr|^{-1}\right) = 2[\tau\gamma(\tau)]^{-1}e^{\tau/2T_1}$. Combining all factors, one arrives at the flux resolution
\begin{equation}
[\delta\Phi] = \Bigl| \frac{d\omega_{01}(\Phi)}{d\Phi} \Bigr|^{-1}
\delta[\Delta\omega(\Phi)] = \Bigl| \frac{d\omega_{01}(\Phi)}{d\Phi} \Bigr|^{-1}
\frac{e^{\tau/2T_1}}{\tau\gamma(\tau)\sqrt{N}}.
\label{eq:nRamsey}
\end{equation}
The standard (classical) measurement is done at a minimal effective delay
$\tau=\tau_0 \ll T_2$. Assuming that each Ramsey experiment takes a time
$T_\mathrm{rep}$, the flux resolution of the standard scheme is given by
Eq.~(\ref{eq:cl_precision}) where $t = NT_\mathrm{rep}$. In a quantum limited measurement, one optimizes the time delay $\tau$. Minimizing the time factor $[\tau\gamma(\tau)]^{-1} e^{\tau/2T_1}$ in Eq.\ (\ref{eq:nRamsey}), one finds the optimal time delay $\tau^*$ from an equation $(2T_1)^{-1} - (\ln[\gamma(\tau)])'=\tau^{-1}$. Considering the $1/f$ flux noise dephasing model (see Methods: Dephasing mechanisms) with $\gamma(\tau) = \exp[-(\Gamma_\mathrm{1/f}\tau)^2]$, we obtain
\begin{equation}
\tau^* = \frac14 \Gamma_\mathrm{1/f}^{-1}\, \Bigl(\sqrt{8
	+(2T_1 \Gamma_\mathrm{1/f})^{-2}} - (2T_1\Gamma_\mathrm{1/f})^{-1} \Bigr).
\end{equation}
As suggested in ref.\ \cite{Bialczak:2007} the $1/f$ flux-noise
originates from spin flips of magnetic impurities located nearby the SQUID
loop. The noise strength then increases linearly with the loop size $L$ giving $\Gamma_\mathrm{1/f} \propto \sqrt{L}$. Hence at large $L$ one has
$\tau^* \to \Gamma_\mathrm{1/f}^{-1}/\sqrt{2} \propto 1/\sqrt{L}$, which
degrades the attainable flux resolution $[\delta\Phi] \propto 1/\tau^*$. The corresponding magnetic field resolution $[\delta B] = [\delta \Phi]/L^2
\propto L^{-3/2}$ still improves with increasing loop size.

{\bf Voltage-to-flux conversion}

The magnetic flux threading the transmon SQUID loop is generated by a
dc-current flowing through the flux-bias line located nearby the SQUID loop
with the current controlled by a dc-voltage $V\in [0.977,1.009]$ V generated with an Agilent 33500B waveform generator (see \SI 2).  As a result, our
device can also be operated as a sensitive voltmeter. The conversion from
voltage values to the non-integer part of the normalized flux
$\left(\Phi/\Phi_0 - n\right)$, where $n$ is an integer number, is obtained
from spectroscopic measurements (Fig.\ 1a), and has the form
\begin{equation}
\left(\frac{\Phi(V)}{\Phi_0} - n\right)
=  \frac{V}{V_0} + \frac{\Phi_{\textrm{tr}}}{\Phi_0}.
\label{eq:flux_voltage_conv}
\end{equation}
Here, $V_0$ is the periodicity (in volts) of the  CPW resonator and qubit
spectra, which corresponds to the magnetic flux change by one flux quantum,
and $\Phi_{\textrm{tr}}$ is the residual flux trapped in the SQUID loop.
Measuring the CPW resonator spectrum periodicity (see Fig.\ 1a
inset), one finds $V_0 = (12.55\pm0.05)$ volts, and the trapped flux value can be found from the position of the qubit spectrum maximum $\omega_{01}
[\Phi(V)]$ (Fig.\ 1a), which gives $\Phi_{\textrm{tr}}/\Phi_0 =
0.059 \pm 0.004$. Hence, our qubit based magnetic flux sensor measures a flux change relative to some reference value.

\vspace{1cm}
{\bf Data availability}

The data that support the findings of this study are available from the
corresponding author upon reasonable request.

\vspace{1cm}
{\bf Acknowledgements}

The work was supported by the Government of the Russian Federation (Agreement 05.Y09.21.0018) (G. L.), by the RFBR Grant No.
17-02-00396A (G. L.), by the Foundation for the Advancement of Theoretical Physics BASIS (G. L.), by the Pauli Center for Theoretical Studies at ETH
Zurich (G. L.) and by the Swiss National Foundation through the NCCR QSIT (A. L.) and the Ministry of Education and Science of the Russian Federation
16.7162.2017/8.9 (A. L.). We acknowledge financial support from V\"aisal\"a
Foundation, the Academy of Finland (project 263457, Centers of Excellence "Low Temperature Quantum Phenomena and Devices" -- project 250280 and "Quantum Technologies Finland" - project 312296), and the Centre for Quantum Engineering at Aalto University.  The experiments used the cryogenic facilities of the Low Temperature Laboratory at Aalto University.
\vspace{1cm}

{\bf Competing interests}

The authors declare that they have no competing interests.

\vspace{1cm}
{\bf Author contribution}

The sample used in the experiment was designed and fabricated by SD. The measurements were performed by SD, AV, and AVL; the data was proceesed by AV and SD and further analyzed by AVL. GBL, GB, and AVL worked on the theoretical foundation of the experiments. AVL and SD wrote the first draft, with further contributions from GBL, GB, and GSP. AVL and GSP supervised the project.

\vspace{1cm}
{\bf References}
\begingroup
\renewcommand{\section}[2]{}

\endgroup

\end{document}